\documentclass[preprint,12pt]{elsarticle}

\usepackage{url}

\usepackage{graphics}

\usepackage{amssymb}

\journal{Physics Letters A}

\begin{document}

\begin{frontmatter}



\title{Phase Shift in the Whitham Zone for the Gurevich-Pitaevskii Special Solution of the Korteweg-de Vries Equation}


\author[im]{R. Garifullin\corref{cor1}}\ead{rustem@matem.anrb.ru}
\author[im]{B. Suleimanov}\ead{bisul@mail.ru}
\author[pot]{N. Tarkhanov}\ead{tarkhanov@math.uni-potsdam.de}

\cortext[cor1]{Corresponding author}

\address[im]{Institute of Mathematics, RAS, 45008, 112, Chernushevskogo, Ufa, Russia}
\address[pot]{Institute of Mathematics, Potsdam University, Am Neuen Palace, 14469, Potsdam, Germany}

\begin{abstract}
We get the leading term of the Gurevich-Pitaevskii special solution of the
KdV equation in the oscillation zone without using averaging methods.

\end{abstract}


\end{frontmatter}





\section{Introduction}
\label{s.Int}
\setcounter{equation}{0}

The Gurevich-Pitaevskii (GP) special universal solution of the Korteweg-de Vries
(KdV) equation
\begin{equation}
\label{eq.KdV}
   u_t + u\, u_x + u_{xxx} = 0
\end{equation}
was introduced in \cite{GurePita71} in connection with the problem of
description of collisionless shock waves
(Sagdeev showed in \cite{Sagd64} that such waves are of oscillating character).
The behaviour of the GP special solution for
   $t \to - \infty$ and
   $x \to \pm \infty$
is determined in the main from the cubic canonical equation of the cusp catastrophe
\begin{equation}
\label{eq.assembly}
   x - t u + u^3 = 0.
\end{equation}

The GP solution to the KdV equation is one of the most interesting special
functions of the modern nonlinear mathematical physics.

In \cite{GurePita71} it is shown that in problems of dispersion hydrodynamics
(in particular, in problems of plasma theory) the GP special solution appears
near the points of overturning of simple waves.
From the results of \cite{KudaSule96a},
                    \cite{KudaSule96b},
                    \cite{KudaSule01}
one actually sees that the same universal special function appears near the
points of overturning of the generic state solutions to diverse dispersion
perturbations of the equations of one-dimensional motion of ideal
incompressible liquid
$$
   \begin{array}{rcl}
     \rho'_t + (v \rho)'_x
   & =
   & 0,
\\
     v'_t + v\, v'_x + \alpha (\rho)\, \rho'_x
   & =
   & 0.
   \end{array}
$$
Here,
   $\rho$ is the density of the liquid,
   $v$ the velocity
and
   $\alpha (\rho) = (c (\rho))^2 / \rho$,
where
   $c (\rho) = \sqrt{p' (\rho)}$ is the speed of sound
and
   $p (\rho)$ the pressure.
In particular, this is the case for solutions of the shallow water equations
$$
   \begin{array}{rcl}
     h'_t + (h A'_x)'_x
   & =
   & \varepsilon^2\, (h^3 A''_{xx})''_{xx} + O (\varepsilon^4),
\\
     \displaystyle
     A'_t + \frac{1}{2}\, (A'_x)^2 + g\, h
   & =
   & \displaystyle
     \frac{1}{2}\, \varepsilon^2 (A'''_{xxt} + A'_x A'''_{xxx} - (A''_{xx})^2)
   + O (\varepsilon^4),
   \end{array}
$$
where
   $h$ is the free boundary,
   $A$ the potential of bottom velocity
and
   $g$ the acceleration of gravity.
The right-hand sides can actually be written as complete series in powers of
the parameter $\varepsilon$ by the procedure given for instance in
   \cite[Ch.~1, \S 4]{Ovsy85}
(and not only as the so-called second approximations, as stated in
   \cite{KudaSule96b}).

In the 1990s there were discovered surprising connections of the GP special
solutions with some problems of quantum gravity.
In \cite{Sule94} this solution was showed to simultaneously satisfy the fourth
order ordinary differential equation
\begin{equation}
\label{eq.ODE}
   u_{xxxx}
 + \frac{5}{3} u u_{xx}
 + \frac{5}{6} (u_x)^2
 + \frac{5}{18}\, (x - t u + u^3)
 = 0,
\end{equation}
which had been studied for $t = 0$ in \cite{BresMariPari90} and
                                      \cite{Moore90}
in connection with evaluating nonperturbative string effects in two-dimensional
quantum gravity
(the equation (\ref{eq.ODE}) belongs to a class of massive string equations).
In \cite{DougSeibShen90} the solution of
$$
   \Big( W^3 - W W_{XX} - \frac{1}{2} (W_X)^2 + \frac{1}{10} W_{XXXX}
   \Big)
 + \frac{15}{32} T \Big( W^2 - \frac13W_{XX} \Big)
 = X
$$
with asymptotics $\sqrt[3]{X}$ as $X \to \pm \infty$ was treated numerically
in connection with problems of quantum gravity.
One can show that this solution $U (t,X)$ is also equivalent to the GP special
solution of (\ref{eq.KdV}) for $t \geq 0$ (but not for $t < 0$).

Dubrovin showed in \cite{Dubr06} and
                        \cite{Dubr08}
directly by means of the theory of approximate symmetries
   \cite{BaikGaziIbra89}
that it is the solution of (\ref{eq.ODE}) with
               asymptotics (\ref{eq.assembly})
that appears near the points of wave overturning for the very diverse singular
dispersion perturbations of the equations of one-dimensional hydrodynamics.

The results of numerical simulations presented in \cite{DougSeibShen90}
demonstrate rather strikingly that the GP special solution of the KdV equation
possesses a domain of undamped oscillations for $t$ large enough.
The authors of \cite{DougSeibShen90} did not conjecture any relation of their
paper to the GP special solution and raised the problem of describing this
domain of oscillations.
Meanwhile Gurevich and Pitaevskii \cite{GurePita74} had used successfully the
self-similar solutions of the averaged Whitham equations \cite{Whit65} to
solve the problem.

The self-similar solutions in question were constructed in explicit form by
Potemin \cite{Pote88}.
However, the problem on the leading term of asymptotics of the GP special
solution in the domain of Whitham oscillations has been open up to now.
One not simple question still unanswered has been that on the phase shift.

Our purpose is to show how it is possible to construct the leading term of the
GP special solution in the zone of oscillations without using any averaging
methods.
To this end we derive certain algebraic equations for
   the slowly varying amplitude and
   the leading term of the phase,
which are actually equivalent to those of \cite{Pote88}.
Moreover, we determine the phase shift of the solution in the oscillation zone.

Our approach may also be of use for the study of undamped oscillations of
other common solutions to integrable partial and ordinary differential
equations which are of importance in physics.
In particular, it applies to two universal solutions of the KdV equation
treated in the recent article \cite{GariSule09}.
Almost one problem in the approach is some awkwardness of analytical
calculations.
However, invoking modern programs for symbol calculations
   (in this paper we use Maple)
often allows one to get rid of such problems without particular difficulties.

\section{Evaluation of phase shift}
\label{s.eopl}
\setcounter{equation}{0}

Consider the solution of the KdV equation that, for
   $t \! \to \! - \infty$ and
   $x \! \to \! \pm \infty$,
is determined in the main from the cubic equation (\ref{eq.assembly}).
It is known that for this solution for positive $t$ there is a domain where
   dissipationless shock waves appear.

We are aimed at constructing asymptotics of the solution in this domain, when
$t \to \infty$.
Following familiar techniques, we change the variables by
\begin{eqnarray*}
   u
 & = &
   \sqrt{|t|}\, U (t,z),
\\
   z
 & = &
   \frac{x}{|t|^{3/2}}.
\end{eqnarray*}
Then equations (\ref{eq.KdV}) and
               (\ref{eq.ODE})
take the form
\begin{equation}
\label{eq.fcov}
\begin{array}{rcl}
   \displaystyle
   t U_t + \frac{1}{2}\, (U - 3 z U_z) + U U_z + t^{-7/2} U_{zzz}
 & =
 & 0,
\\
   \displaystyle
   t^{-7} U_{zzzz}
 + \frac{5}{18}\, t^{-7/2} (6 U U_{zz} + 3 (U_z)^2)
 + \frac{5}{18}\, (z - U + U^3)
 & =
 & 0.
\end{array}
\end{equation}

We now look for a solution $U$ of the system in the form of asymptotic series
\begin{equation}
\label{eq.asymptotic}
   U
 = U_0 (\varphi,z)
 + t^{- 7/4} U_1 (\varphi,z)
 + t^{- 7/2} U_2 (\varphi,z)+\ldots,
\end{equation}
where $U_0$,
      $U_1$ and
      $U_2$
are $2 \pi\,$-periodic in the fast variable $\varphi$.
This latter is assumed to be of the form
$$
   \varphi = t^{- 7/4} f (z) + s (z),
$$
where by $s (z)$ is meant precisely the phase shift.

For the unknown function $U_0$ we get the nonlinear system
$$
\begin{array}{rcl}
   Q^3 \partial_\varphi^3 U_0
 + Q R \partial_\varphi U_0
 + Q U_0 \partial_\varphi U_0
 & =
 & 0,
\\
   \displaystyle
   Q^4 \partial_\varphi^4 U_0
 + \frac{5}{6}\, Q^2 (2 U_0 \partial_\varphi^2 U_0 + (\partial_\varphi U_0)^2)
 + \frac{5}{18}\, (z - U_0 + U_0^3)
 & =
 & 0,
\end{array}
$$
while the systems for $U_1$
$$
\begin{array}{rcl}
   Q^3 \partial_\varphi^3 U_1
 + Q (R + U_0) \partial_\varphi U_1
 + Q \partial_\varphi U_0\, U_1
 & \! \! = \! \!
 & F_1,
\\
   \displaystyle
   Q^4 \partial_\varphi^4 U_1
 + \frac{5}{3}\, Q^2
   (U_0 \partial_\varphi^2 U_1 + \partial_\varphi U_0 \partial_\varphi U_1)
 + \frac{5}{18}\, (3 U_0^2 + 3 Q^2 \partial_\varphi^2 U_0 - 1) U_1
 & \! \! = \! \!
 & F_2,
\end{array}
$$
and for $U_2$
$$
\begin{array}{rcl}
   Q^3 \partial_\varphi^3 U_2
 + Q (R + U_0) \partial_\varphi U_2
 + Q \partial_\varphi U_0\, U_2
 & \! \! = \! \!
 & G_1,
\\
   \displaystyle
   Q^4 \partial_\varphi^4 U_2
 + \frac{5}{3}\, Q^2
   (U_0 \partial_\varphi^2 U_2 + \partial_\varphi U_0 \partial_\varphi U_2)
 + \frac{5}{18}\, (3 U_0^2 + 3 Q^2 \partial_\varphi^2 U_0 - 1) U_2
 & \! \! = \! \!
 & G_2
\end{array}
$$
proves to be linear.
Here, $F_1$ and
      $F_2$ are explicit functions depending on $z$ and $U_0$,
and $G_1$ and
    $G_2$ are explicit functions depending on $z$ and $U_0$, $U_1$,
i.e.,
   the right-hand sides are explicit functions depending on $z$ and on the
   preceding corrections.
We write
\begin{equation}
\label{eq.Q,R}
\begin{array}{rcl}
   Q & = & f',
\\
   R & = & \displaystyle \frac{7}{4}\, \frac{f}{f'} - \frac{3}{2} z
\end{array}
\end{equation}
for short.

From the compatibility condition of the equations for $U_0$ we obtain a first
order equation
\begin{equation}
\label{eq.U0}
   Q^2 (\partial_\varphi U_0)^2
 + \frac{1}{3}\, U_0^3
 + R\, U_0^2
 + \frac{1}{3}\, (18 R^2 - 5) U_0
 + \frac{1}{3}\, (15 R - 54 R^3 - 5 z)
 = 0.
\end{equation}
From the compatibility condition of the equations for $U_1$ we derive a
nonlinear equation
\begin{equation}
\label{eq.U1}
   \frac{d}{dz} R
 = \frac{1}{9}\,
   \frac{486\, R^4 - 171\, R^2 + 9 z R + 5}
        {(54\, R^3 - 9 R + z) (2 R + 3 z)}
\end{equation}
for the unknown function $R = R (z)$.
(In Section \ref{s.roetsf} we show that this equation agrees with results
 obtained earlier.)
When requiring the compatibility of the equations for $U_2$, we deduce that
the function $U_0 (\varphi,z)$ should satisfy, together with (\ref{eq.U0}),
a nonlinear ordinary differential equation in the variable $z$ of the form
\begin{equation}
\label{eq.U2}
   \partial_z^2 U_0
 - \frac{\partial_{\varphi}^2 U_0}{(\partial_{\varphi} U_0)^2}\,
   (\partial_z U_0)^2
 + \frac{P_3 (U_0)}{(\partial_{\varphi} U_0)^2}\, \partial_z U_0
 + \partial_{\varphi} U_0\, (s'' + H s')
 + \frac{P_4 (U_0)}{(\partial_{\varphi} U_0)^2}
 = 0.
\end{equation}
Here,
   $P_3 (U_0)$ and
   $P_4 (U_0)$
are polynomials in $U_0$ of degrees $3$ and
                                    $4$,
respectively, with coefficients depending on $z$ and
                                             $R$.
The function $H = H (z,R)$ is given by
$$
   H(z,R)
 = \frac{1}{3}\,
   \frac{N(z,R)}
        {(54\, R^3 - 9 R + z)^2 (2 R + 3 z)^2}
$$
where
$$
\begin{array}{rcl}
   N(z,R)
 \! & \! = \! & \!
   45 z^3
 +
   (4860 R^3 \! - \! 582 R) z^2
 +
   (131220 R^6 \! - \! 43416 R^4 \! + \! 2721 R^2 \! - \! 35) z
\\
 \! & \! + \! & \!
   139968 R^7 \! - \! 59616 R^5 \! + \! 6048 R^3 \! - \! 120 R.
\end{array}
$$

Note that (\ref{eq.U2}) is a Hamiltonian equation with Hamiltonian quadratic
relative to the impulse, i.e.
   $a (\partial_z U_0)^2 + b \partial_z U_0 + c$
where $a$, $b$ and $c$ are functions of $z$ and $U_0$.

We proceed to study equations (\ref{eq.U0})-(\ref{eq.U2}).
Equation (\ref{eq.U0}) is autonomous in the fast variable $\varphi$, hence the
arbitrary constant of the general solution is contained in the phase shift
which we take into account in the variable $s (z)$.
The general solution of this equation is sought in the form
$$
   U_0 = A\, \mathrm{dn}^2 \Big( \frac{B}{Q} \varphi; k \Big) + C,
$$
where
   $\mathrm{dn}$ is the elliptic function of Jacobi
and
   $A$, $B$, $C$ and $k$ are to be defined.
On substituting $U_0$ into equation (\ref{eq.U0}) and
   equating the coefficients of different powers of the Jacobi function to
   zero
we get the system of algebraic equations
\begin{equation}
\label{eq.algebraic}
   \begin{array}{rcl}
     A^2 (A - 12 B^2)
   & =
   & 0,
\\
     A^2 ((8 - 2 k^2) B^2 + C + R)
   & =
   & 0,
\\
     A\, ((12 k^2 - 12)  A B^2 + 18 A^2 + 6 R C - 5)
   & =
   & 0,
\\
     C^3 + 3 R C^2 + (18 R^2 - 5) C - 54 R^3 + 15 R - 5 z
   & =
   & 0.
   \end{array}
\end{equation}
From the assumption on the $2 \pi\,$-periodicity of $U_0$ it follows that
\begin{equation}
\label{eq.periodicity}
   \frac{B}{Q} = \frac{K (k)}{\pi},
\end{equation}
where
   $K (k)$ is the complete elliptic integral of first kind.

The system of equations (\ref{eq.Q,R}),
                        (\ref{eq.U1}),
                        (\ref{eq.algebraic}) and
                        (\ref{eq.periodicity})
obtained in this way is overdetermined.
It consists of 6 algebraic equations and
               2 differential equations
for the unknowns $f$, $Q$, $R$, $A$, $B$, $C$ and $k$.
Our next concern will be to show that one can find all slowly varying unknowns
without solving the differential equations.

The unknowns $f$, $Q$, $R$, $B$ and $C$ can be determined immediately from
this system through $A$, $k$ and $z$.
More precisely,
\begin{equation}
\label{eq.f,Q,R,B,C}
\begin{array}{rclrcl}
   B
 & \! \! \! = \! \! \!
 & \displaystyle
   \sqrt{\frac{A}{12}},
 & C
 & \! \! \! = \! \! \!
 & \displaystyle
   \frac{(k^2\!-\!2) (4 k^4\!-\!5 k^2\!+\!5) A^3\!-\!10 (K (k)\!-\!2) A\!-\!45 z}
        {14 (k^4 - k^2 + 1) A^2 - 30},
\\
   R
 & \! \! \! = \! \! \!
 & \displaystyle
   \frac{(k^2\!-\!2) A}{3}\!-\!C,
 & Q
 & \! \! \! = \! \! \!
 & \displaystyle
   \frac{\pi B}{K (k)},
\\
   f
 & \! \! \! = \! \! \!
 & \displaystyle
   \frac{Q (4 R\!+\!6 z)}{7}.
 &
 &
 &
\end{array}
\end{equation}
On substituting these expressions into (\ref{eq.algebraic}) we arrive at one
algebraic equation
\begin{equation}
\label{eq.A,k}
\begin{array}{rcl}
 & \! \! \! \! \! \! &
   (8 k^{12}\!-\!24 k^{10}\!+\!43 k^8\!-\!46 k^6\!-\!43 k^4\!+\!24 k^2\!-\!8)\,
   A^6
 \! - \!
   140 (k^4\!-\!k^2\!+\!1)^2\, A^4
\\
 & \! \! \! + \! \! \! &
   50 z (k^2\!-\!2) (2 k^2\!-\!1) (k^2\!+\!1)\, A^3
 \! + \!
   500 (k^4\!-\!k^2\!+\!1)\, A^2
 \! + \!
   3375 z^2\!=\!500
\end{array}
\end{equation}
for $A$ and $k$.
Differentiating this equality in $z$ and
substituting the resulting expression along with (\ref{eq.f,Q,R,B,C}) into
   (\ref{eq.Q,R}) and
   (\ref{eq.U1}),
we obtain 3 equations containing $A' (z)$ and $k' (z)$.
On eliminating these derivatives, we get another algebraic equation for $A$
                                                                    and $k$
which contains the quotient of two complete elliptic integrals
$$
   q = \frac{E (k)}{K (k)}.
$$
Using (\ref{eq.A,k}) to eliminate the highest power of $z$ from the latter
equation we bring it to the form
\begin{equation}
\label{eq.k,A}
   \begin{array}{rcl}
 & \! \! \! \! \! \! &
   21 k^4 (k^2 \! - \! 1)^2 A^3
 \! + \!
   10\,
   ((2 q \! - \! 1) k^6 \!
 - \! (3 q \! + \! 1) k^4 \!
 - \! (3 q \! - \! 4) k^2 \!
 + \! (2 q \! - \! 2)) A
\\
 & \! \! \! + \! \! \! &
   315 z\, ((2 q \! - \! 1) k^4 \!
          - \! (2 q \! - \! 3) k^2 \!
          + \! (2 q \! - \! 2))\, =\, 0.
   \end{array}
\end{equation}
When eliminating the variable $A$ from (\ref{eq.A,k}) and
                                       (\ref{eq.k,A}),
one finds $k$ as an implicit function of $z$.
The other functions can be expressed explicitly through $k$.

This method allows one to get the explicit formulas of \cite{Pote88} without
using averaging procedure.

The domain of $z$ in which oscillations are possible is determined from the
condition that $k \in [0,1]$.
To the point $k = 0$ there corresponds leading wave front set
   $z_l = - \sqrt{2}$
and to the point $k = 1$ there corresponds trailing wave front set
   $z_t = \sqrt{10}/27$.

The second ordinary differential equation of (\ref{eq.fcov}) enables us also
to determine the phase shift $s (z)$ of the solution.
For this purpose we make use of equation (\ref{eq.U2}).
Note that $U_0$ is an even function of $\varphi$, hence
   $\partial_{\varphi} U_0$ is odd
and
   $\partial_z U_0$,
   $\partial_z^2 U_0$ are even
in $\varphi$.
Multiplying (\ref{eq.U2}) by $(\partial_{\varphi} U_0)^3$,
integrating in $\varphi$ over the whole period $2 \pi$ and
taking into account that the mean value of an odd periodic function over the
whole period vanishes, we get the equation
\begin{equation}
\label{eq.shift}
   s'' + H (z,R) s' = 0.
\end{equation}

In order to choose a concrete solution to (\ref{eq.shift}), one has to use the
asymptotics of the solution for $z \to z_l$.
Such an asymptotics is given in \cite{KudaSule96a} and it shows in particular
that the solution does not contain $\log (z-z_l)$.
In our case we get
\begin{eqnarray*}
   k
 \! & \! \! = \! \! & \!
   \frac{2^{7/8}}{\sqrt{5}}
   (z \!\! - \!\! z_l)^{1/4}
   \Big( 1
       \! - \! \frac{2^{3/4}}{10} (z \!\! - \!\! z_l)^{1/2}
       \! + \! \frac{131}{1280} \sqrt{2}\, (z \!\! - \!\! z_l)
       \! + \! O ((z \!\! - \!\! z_l)^{3/2})
   \Big),
\\
   a
 \! & \! \! = \! \! & \!
 - \frac{\sqrt{2}}{6}
 + \frac{1}{40}\, (z \!\! - \!\! z_l)
 + \frac{7}{2560} \sqrt{2}\, (z \!\! - \!\! z_l)^2
 + O ((z \!\! - \!\! z_l)^3),
\\
   H
 \! & \! \! = \! \! & \!
   \frac{1}{z \!\! - \!\! z_l}
 - \frac{543}{1600} \sqrt{2}
 + O (z \!\! - \!\! z_l).
\end{eqnarray*}

From the asymptotics of $H$ it follows that a solution of (\ref{eq.shift})
contains terms of the form $\log (z + \sqrt{2})$ which are not permitted by
\cite{KudaSule96a}.
Hence, this solution enters into the linear combination of solutions with
coefficient $0$, and so
$$
   s (z) = s_0
$$
is constant.

The hypothesis on the constancy of the phase shift $s (z)$ has been formulated
in \cite{KudaSule96a}.
However, in \cite{KudaSule96a} it was based solely on the asymptotics given
there.
It is clear that {\it a priori} one might not exclude the situation where the
phase shift fails to be constant but tends exponentially fast to a constant
as $z \to z_l$.
From (\ref{eq.shift}) and the asymptotics of $H (z,R)$ for $z \to z_l$ we see
that such is not the case, and so $s (z)$ proves to be constant.

To evaluate the constant $s_0$ we invoke a numerical simulation.
Namely, we compare a numerical solution with the solution constructed by using
asymptotic formulas.

\section{Numerical simulations}
\label{s.ns}
\setcounter{equation}{0}

\begin{figure}[htb]
\includegraphics [width=12.8cm,keepaspectratio] {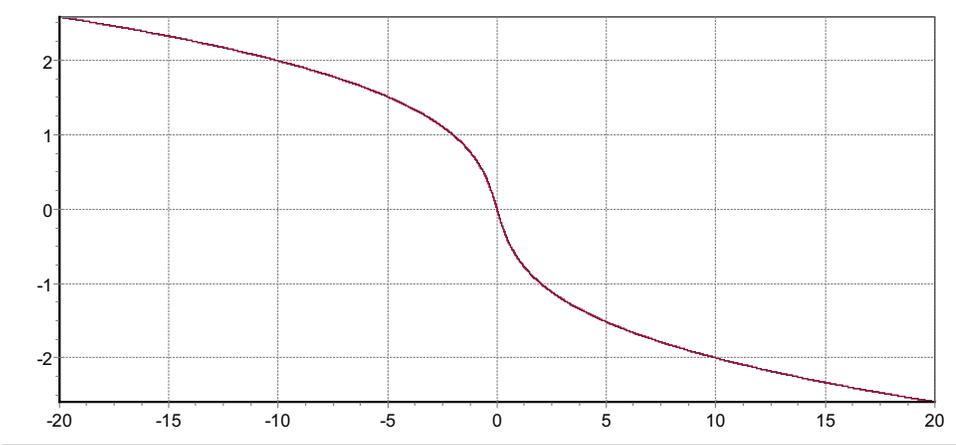}
\caption{%
The numerical simulation for the function $U(t,z)$ corresponding to
$t = -7$.
}
\label{f.0}
\end{figure}

\begin{figure}[htb]
\includegraphics [width=12.8cm,keepaspectratio] {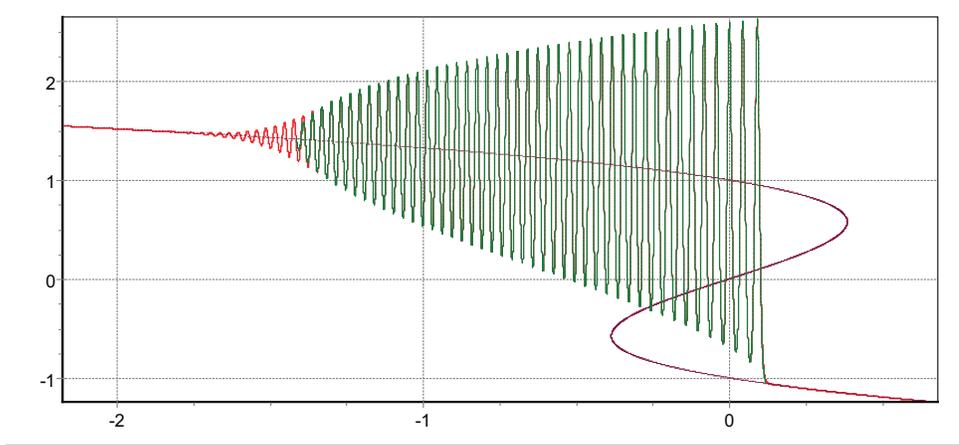}
\caption{%
The numerical simulation for the function $U(t,z)$ corresponding to
$t = 20$ and root of equation $z-\Lambda+\Lambda^3=0$.%
}
\label{f.1}
\end{figure}

\begin{figure}[htb]
\includegraphics [width=10.8cm,keepaspectratio] {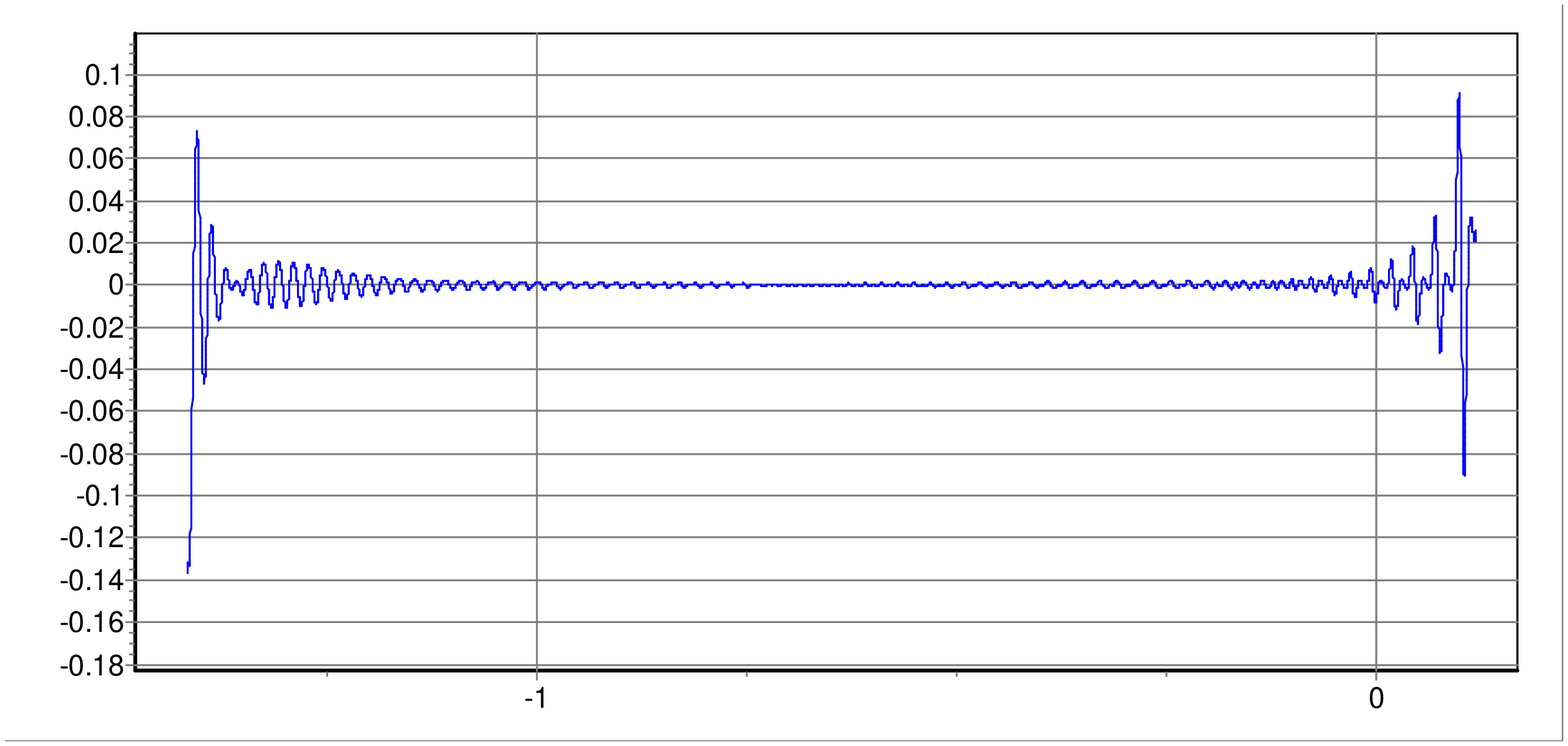}
\caption{%
The difference between the numerical solutions and asymptotic for $U(t,z)$ for $t = 20$.%
}
\label{f.2}
\end{figure}

To this end we have written a special program.
The results of numerical simulations are presented in Figures \ref{f.0} --
                                                             \ref{f.2}.

In Figures~\ref{f.0}-\ref{f.1} one can observe numerical solutions for function $U(t,z)$ for negative and positive value $t$. It can be shown that function $U(t,z)$ for negative value $t$ practically councide with the root of the equation $\Lambda^3+\Lambda=-z$.

In Figure~\ref{f.2} the difference between these solutions is shown.
Both figures correspond to $t = 20$.
The constant $s_0$ has proved to be equal to $3.1254 \approx \pi$.
The difference between two solutions is a multiple of
   $t^{-1.77} \approx t^{-7/4}$
which agrees with the order of the first correction in formula
   (\ref{eq.asymptotic}).

We believe that the constant $s_0$ just amounts to $\pi$ and the small
difference is caused by a computation error.

Figure~\ref{f.2} makes it evident that the error increases for $z$ close to
$- \sqrt{2}$ and
$\sqrt{10} / 27$.
This manifests the nonuniform character of the constructed asymptotic formula
in the entire domain of the variable $z$.
In neighbourhoods of the leading and
                         trailing
wave front sets one should construct other asymptotic formulas which can be
made consistent with asymptotic expansion (\ref{eq.asymptotic}),
   as it is described in \cite{Il'in92}.

\section{Reduction of equations to standard form}
\label{s.roetsf}
\setcounter{equation}{0}

To start with, we bring the equations obtained in \cite{Pote88} to a simpler
form.
These are
$$
   z - Y_j + Z_j = 0
$$
for $j = 1, 2, 3$,
   where
$$
\begin{array}{rclrclrcl}
   Y_1
 & \! \! = \! \!
 & \displaystyle
   \frac{1}{3} S_1 \! + \! \frac{2}{3} \frac{l_1\!-\!l_2}{1\!-\!q},
 & Y_2
 & \! \! = \! \!
 & \displaystyle
   \frac{1}{3} S_1 \! - \! \frac{2}{3} \frac{(l_1\!-\!l_2) (1\!-\!k^2)}
                                            {1\!-\!q\!-\!k^2},
 & Y_3
 & \! \! = \! \!
 & \displaystyle
   \frac{1}{3} S_1 \! + \! \frac{2}{3} \frac{l_3\!-\!l_2}{q},
\\
   S_1
 & \! \! = \! \!
 & l_1 + l_2 + l_3,
 & S_2
 & \! \! = \! \!
 & l_1 l_2 + l_2 l_3 + l_3 l_2,
 & S_3
 & \! \! = \! \!
 & l_1 l_2 l_3
\\
\end{array}
$$
and
$$
   Z_j = \frac{1}{35} \Big( V + (3 Y_j - S_j) V'_{l_j} \Big)
$$
with $V = 5 S_1^3 - 12 S_1 S_2 + 8 S_3$.

Obviously, one can eliminate two variables from three equations.
On eliminating $z$ and $q$ we get an equation for $l_1$, $l_2$ and $l_3$,
which does not contain $z$ and $q$.
Namely,
\begin{equation}
\label{eq.first}
   3\, (l_1^2 + l_2^2 + l_3^2)
 + 2\, (l_1 l_2 + l_2 l_3 + l_3 l_1)
 - 5
 = 0.
\end{equation}

We now eliminate either of the variables $z$ and
                                         $q$
in the equations and use (\ref{eq.first}) to eliminate all powers of $l_1$
greater than the first one.
Then we get two more equations
\begin{eqnarray}
\label{eq.second}
   z
 & = &
   \frac{2}{45}
   \Big( l_1 (8 l_2^2 + 4 l_2 l_3 + 8 l_3^2 - 15)
      - (l_2 + l_3) (24 l_2^2 - 8 l_2 l_3 + 24 l_3^2 - 25)
   \Big),
\nonumber
\\
   q
 & = &
   \frac{1}{2}\,
   \frac{(l_2 - l_3) (3 l_2 l_3 + 3 l_3 l_1 + 9 l_3^2 - 5)}
        {l_1 (2 l_2^2 + l_2 l_3 + 2 l_3^2 - 5)
         - (l_2 + l_3) (6 l_2^2 - 2 l_2 l_3 + 6 l_3^2 - 5)}.
\nonumber
\\
\end{eqnarray}

The system of equations (\ref{eq.Q,R}),
                        (\ref{eq.algebraic}) and
                        (\ref{eq.periodicity})
is equivalent to the system of averaged equations (\ref{eq.first}) and
                                                  (\ref{eq.second})
which was obtained in \cite{Pote88} by Whitham's method.
To prove this, we pass in the system of
   the equation for $C$ in (\ref{eq.f,Q,R,B,C}) and equations (\ref{eq.A,k}),
                                                              (\ref{eq.k,A})
to Whitham's variables $l_1 (z)$, $l_2 (z)$ and $l_3 (z)$ by 
$$
\begin{array}{rclrclrcl}
   A
 & \! \! = \! \!
 & 2\, (l_3 - l_1),
 & C
 & \! \! = \! \!
 & l_1 + l_2 - l_3,
 & k^2
 & \! \! = \! \!
 & \displaystyle
   \frac{l_2 - l_1}{l_3 - l_1}.
\end{array}
$$
On eliminating $z$ and $q$ from the obtained equations we arrive precisely at
(\ref{eq.first}).
On eliminating either of $z$ and $q$ and all powers of $l_1$ greater than the
first one, we get (\ref{eq.second}), as desired.

\section{Conclusion}
\label{s.conclusion}
\setcounter{equation}{0}

It should be noted that the value of the phase shift $s_0$ derived from the
numerical simulation contradicts \cite{KudaSule96a,KudaSule96b}.
From the results of the paper it follows that the function $s (z)$ tends to
$\pi / 2$, as $z \to z_l$.
But there is an arithmetical error in formula (17)\cite{KudaSule96a} and in formula (21)\cite{KudaSule96b}. The right calculation with using monodromic date from \cite{Kapaev91}
showed that $s(z)$ tends to $\pi$, and so $$s(z)\equiv\pi.$$

\bigskip

{\bf Acknowledgements\,}
The authors are greatly indepted to V. Adler for first numerical experiments in this problem. 
The derivation of equation (\ref{eq.U1}) without using the averaging method is due to our colleague V. Kudashev\footnote{V. Kudashev died on 1999.}. 
The research of the first author was supported by the DFG grant TA 289/4-1 and by the Program for Supporting Young Scientists, grant MK-2812.2010.1.
The first and second authors were also supported by the RFBR 09-01-92436, 10-01-91222.


\end{document}